\documentclass [ twocolumn, secnumarabic, amssymb, nobibnotes, aps, prl,showpacs,superscriptaddress]{revtex4-1}
\usepackage{graphicx}
\usepackage{float}
\usepackage{bm}

\begin{document}

\title{Doping evolution of the anisotropic upper critical fields in iron-based superconductor Ba$_{1-x}$K$_x$Fe$_2$As$_2$ }

\author{M.~A.~Tanatar}
\affiliation{Ames Laboratory US DOE, Ames, Iowa 50011, USA}
\affiliation{Department of Physics and Astronomy, Iowa State University, Ames, Iowa 50011,
USA }

\author{Yong~Liu}
\affiliation{Ames Laboratory US DOE, Ames, Iowa 50011, USA}

\author{J.~Jaroszynski}
\affiliation{National High Magnetic Field Laboratory, Florida State University, Tallahassee, Florida 32310, USA}

\author{J.~S.~Brooks}
\affiliation{National High Magnetic Field Laboratory, Florida State University, Tallahassee, Florida 32310, USA}

\author{T.~A.~Lograsso}
\affiliation{Ames Laboratory US DOE, Ames, Iowa 50011, USA}
\affiliation{Department of Materials Science and Engineering, Iowa State University, Ames, Iowa 50011, USA }

\author{R.~Prozorov}
\affiliation{Ames Laboratory US DOE, Ames, Iowa 50011, USA}
\affiliation{Department of Physics and Astronomy, Iowa State University, Ames, Iowa 50011,
USA }

\date{\today}

\begin{abstract}
In-plane resistivity measurements as a function of temperature and magnetic field up to 35~T with precise orientation within the crystallographic $ac-$plane were used to study the upper critical field, $H_{c2}$, of the hole-doped iron-based superconductor Ba$_{1-x}$K$_x$Fe$_2$As$_2$. Compositions of the samples studied were spanning from underdoped $x=$0.17 ($T_c$=12~K) and $x$=0.22 ($T_c$=20~K), both in the coexistence range of stripe magnetism and superconductivity, though optimal doping $x$=0.39 ($T_c$=38.4~K), $x$=0.47 ($T_c$=37.2~K), to overdoped $x$=0.65 ($T_c$=22~K), $x$=0.83 ($T_c$=10~K). We find notable doping asymmetry of the shapes of the anisotropic $H_{c2}(T)$  suggesting important role of paramagnetic limiting effects in $H \parallel a$ configuration in overdoped compositions and multi-band effects in underdoped compositions.

\end{abstract}


\maketitle

\section{Introduction}

Distinctive features of the iron-based high transition temperature, $T_c$, superconductors \cite{Hosonoreview} are very high values of the upper critical fields, $H_{c2}$, \cite{JJ} and their low anisotropy with respect to the  Fe-As  layer (tetragonal or orthorhombic $ab$-plane), $\gamma_H = H_{c2}^{ab} /H_{c2}^{c}$ \cite{NiHc2,Altrawneh,Singleton,Yuan2}. 
Anisotropy of the upper critical field in the orbital limiting scenario \cite{WHH} is determined by the anisotropy of the Fermi velocity, and thus is linked with resistivity anisotropy $\gamma_{\rho} \equiv \rho_c/\rho_a$ with $\gamma_{\rho} \approx \gamma_H^2$ at $T_c$ \cite{anisotropy,YLiu1}. In uniaxial (tetragonal and hexagonal) crystals, the dependence of the orbital $H_{c2}$ on angle $\theta$ with respect to the $ab$-plane can be written as :

\begin{equation}
H_{c2}(\theta)=\frac{H_{c2}^{ab}}{\sqrt{(\gamma_H ^2 -1) \sin^2\theta + 1 }}.
\label{Hc2-theta}
\end{equation}

\noindent

Notable deviations from this angular dependence were found in electron over-doped Ba(Fe$_{1-x}$Ni$_x$)$_2$As$_2$ and discussed in multi-band scenario \cite{JasonHc2}. Deviations can be particularly pronounced when magnetic field is aligned parallel to the conducting plane, so that orbital upper critical fields can become higher than paramagnetic limit \cite{CC}. Crossover between the orbital and paramagnetic limiting  mechanisms leads to a difference in the shape of the $H_{c2}(T)$ line, which was noted in KFe$_2$As$_2$ \cite{TerashimaHc2,YLiu1} and nearby hole-overdoped Ba$_{1-x}$K$_x$Fe$_2$As$_2$ compositions \cite{TerashimaBaKhysteresis,almasan}. The importance of the paramagnetic limiting effects was also suggested by the observation of the first order transition at $H_{c2}^{ab}(T)$ line at low temperatures in thermal expansion and magnetostriction measurements in KFe$_2$As$_2$ \cite{Zocco},  small angle neutron scattering \cite{Eskildsen} and anomalous hysteresis in field-sweep resistivity measurements \cite{TerashimaBaKhysteresis}. Since close to $T_c$ $H_{c2}$ is always determined by orbital mechanism, one needs to study low temperatures regime 
where superconducting gap is fully developed.
Thus far low-temperature measurements in the under-doped Ba$_{1-x}$K$_x$Fe$_2$As$_2$ compositions were performed only in $H \parallel c$ configuration \cite{ReidBaKHC2}. 
To the best of our knowledge there were no studies of the anisotropy of $H_{c2}$  in the conducting plane of any of the iron-based superconductors, which is usually neglected as being small compared to $ac$-plane anisotropy in line with experimental studies in some compounds \cite{SrRuOanisotropy,lambdaBETS,CeCoIn5anisotropy}.

In hole doped materials Ba$_{1-x}$K$_x$Fe$_2$As$_2$ the slope of the $H_{c2}^{c}(T)$ curves close to zero-field $T_c$ over a broad composition range 0.22$\leq x \leq 1$  scales well with $T_c$, while the $\gamma_H$ anisotropy somewhat increases for $x>0.83$ close to $x$=1 \cite{YLiu1,yliu2}. Interestingly this is the composition range where the superconducting gap also becomes nodal \cite{Fukazawa,SLi,octet,thermalcondMatsuda,thermalcondLi,ReidK,KChoSA}. In the end hole-doped composition KFe$_2$As$_2$, the upper critical field $H_{c2}^{c}$ strongly changes upon pressure-induced transition \cite{Valentin} between two different superconducting states \cite{FazelNP,FazelPRB,Terashima2states}. It was argued that the transformation is consistent with the transformation of the superconducting gap structure, namely development of horizontal nodes in the superconducting gap \cite{Valentin}. 

The dependence of the upper critical field on the superconducting gap structure, on proximity to magnetism and on the topology of the Fermi surface makes doping evolution of the upper critical field in hole-doped Ba$_{1-x}$K$_x$Fe$_2$As$_2$ non-trivial. The superconducting state of Ba$_{1-x}$K$_x$Fe$_2$As$_2$ (see doping phase diagram in the bottom panel of Fig.~\ref{resistivity})
has ranges of coexistence with two different types of magnetism (stripe antiferromagnetic $C_2$ phase \cite{Rotter} and tetragonal antiferromagnetic $C_4$ phase \cite{Hassinger,Bohmer,C4Avci}). The anisotropy of the superconducting gap notably increases in the $C_2$AF - SC coexistence range \cite{HyunsooBaK,ReidBaKHC2}, similar to overdoped compositions. In addition Fermi surface topology changes at $x \sim$0.5 \cite{HalynaBaK} and $x \sim $0.7-0.8 \cite{Ding}, the latter also being accompanied by the superconducting gap anisotropy change \cite{thermalcondMatsuda,KChoSA}.

In this article we report comparative study of the precision alignment anisotropic $H_{c2}(T)$ for underdoped and overdoped compositions of hole-doped  Ba$_{1-x}$K$_x$Fe$_2$As$_2$ using DC magnetic field up to 35~T in National High Magnetic Field Laboratory in Tallahassee. The compositions were selected with close values of $T_c$ in 10~K range ($x=$0.17, $T_c$=12~K and $x$=0.83, $T_c$=10~K), in 20~K range ($x$=0.22, $T_c$=20~K and $x$=0.65, $T_c$=22~K), and in 38~K range close to optimal doping ($x$=0.39, $T_c$=38.4~K and  $x$=0.47, $T_c$=37.2~K). The whole $H-T$ phase diagram could be explored in 10~K class samples, large part of it in 20~K class samples and only a small range in the optimally doped samples. Our main findings are clear tendency for paramagnetic limiting effects on the overdoped side of the phase diagram and notable difference in  the shape of $H_{c2}^{ab}(T)$ lines in samples of similar $T_c$ in the over-doped and under-doped range.   


\section{Experimental}

\begin{figure}[tb]
\begin{center}
\includegraphics[width=0.90\linewidth]{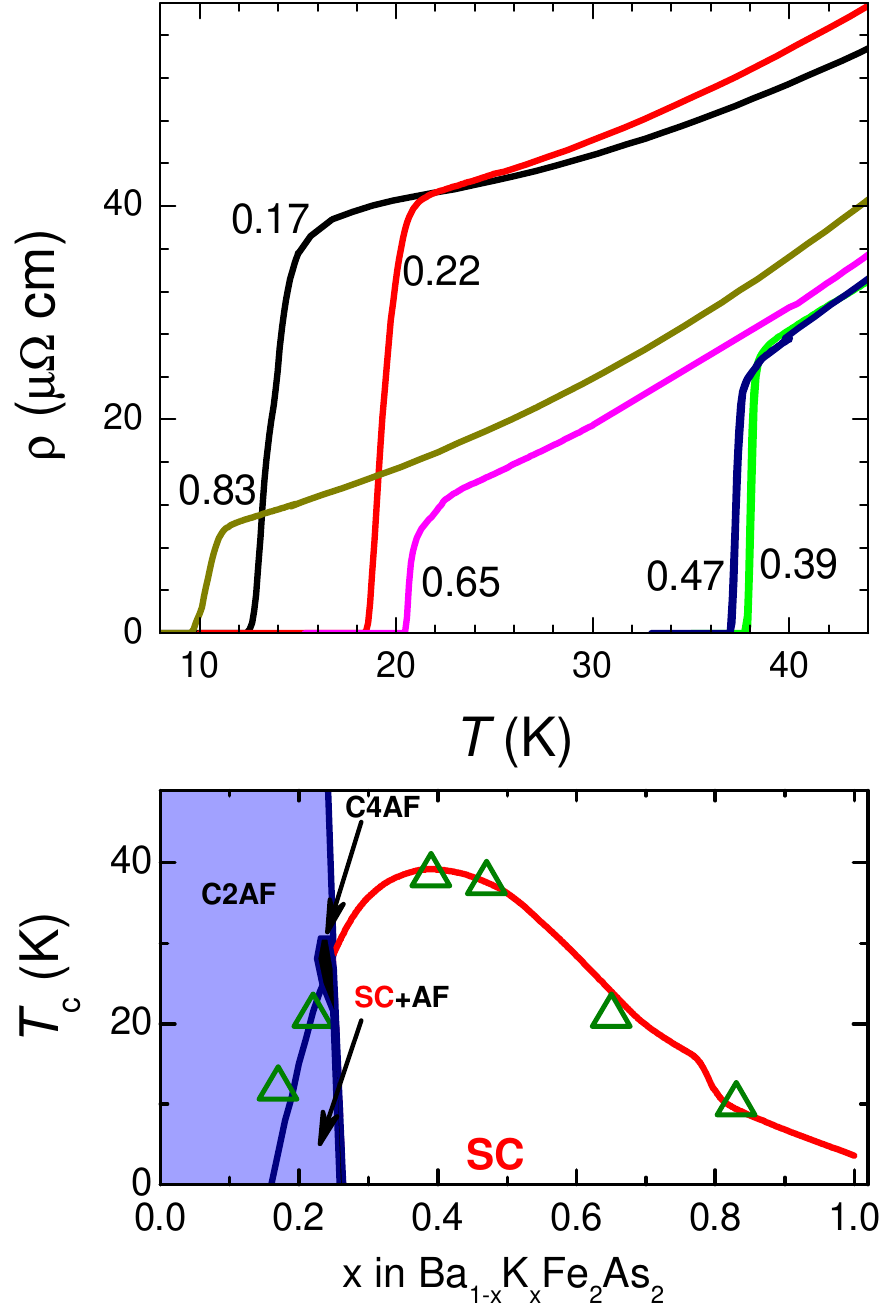}
\end{center}
\caption{(Color Online) (Top panel) Temperature-dependent electrical resistivity of selected representative samples of Ba$_{1-x}$K$_x$Fe$_2$As$_2$ with $x$=0.17, 0.22, 0.39, 0.47, 0.65 and 0.83. Compositions were chosen so that to have $T_c$ of about 10~K on overdoped ($x$=0.83) and underdoped ($x$=0.17) sides, 20~K ($x$=0.65 and 0.22) and above 35~K ($x$=0.39 and 0.47), respectively. Lower panel shows doping phase diagram with position of the samples studied. C2AF corresponds to a range of stripe antiferromagnet, C4AF corresponds to a range of tetragonal $C_4$ antiferromagnetic phase, C4PM corresponds to tetragonal paramagnetic state, SC is domain of superconductivity, including ranges of coexistence with C2AF and C4AF phases (SC+AF).   
}%
\label{resistivity}
\end{figure}

\begin{figure}[tb]
\begin{center}
\includegraphics[width=0.90\linewidth]{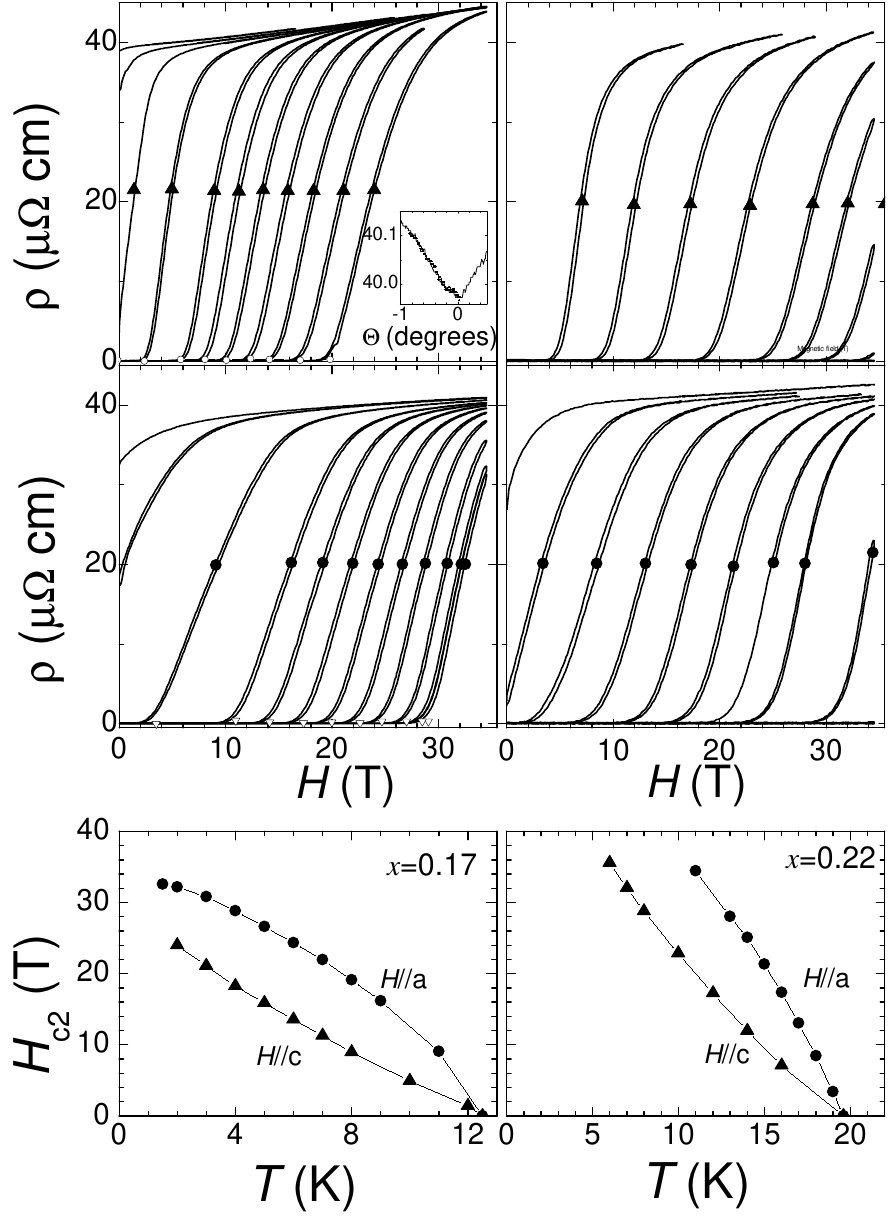}
\end{center}
\caption{(Left column of panels) Magnetic field-dependent resistivity of underdoped sample Ba$_{1-x}$K$_x$Fe$_2$As$_2$ $x$=0.17 taken in isothermal conditions in magnetic fields oriented along tetragonal $c$-axis ($H \parallel c$, top panel, temperatures 16~K, 14~K, 12~K, 10~K, 8~K, 7~K, 6~K, 5~K, 4~K, 3~K and 2~K left to right) and perpendicular to it ($H \parallel a$, mid panel, temperatures from left to right 15~K, 13~K, 11~K, 9~K, 8~K, 7~K, 6~K, 5~K, 4~K, 3~K, 2~K, 1.8~K). Bottom panel shows $H-T$ phase diagrams for two field orientations determined using mid-point criterion between up- and down- field sweeps.  The inset to the top panel shows sample alignment procedure. Resistivity measurements were taken in field $H$ slightly lower than $H_{c2}^{a}$, in which sample resistance shows strong angular dependence. The curve was measured in one-direction motion of the rotator to avoid backlash, with deep minimum corresponding to $H \parallel ab$.   
(Right column of panels.) 
Magnetic field-dependent resistivity of underdoped sample Ba$_{1-x}$K$_x$Fe$_2$As$_2$ $x$=0.22 taken in isothermal conditions in magnetic fields oriented along tetragonal $c$-axis ($H \parallel c$, top panel, temperatures 16~K, 14~K, 12~K, 10~K, 8~K, 7~K, and 6~K left to right) and perpendicular to it ($H \parallel a$, mid panel, temperatures from left to right 20~K, 18~K, 17~K, 16~K, 15~K, 14~K, 13~K, and 11~K). Bottom panel shows $H-T$ phase diagrams for two field orientations determined using mid-point criterion between up- and down- field sweeps.
}%
\label{x0p17andx0p22}
\end{figure}

Single crystals of Ba$_{1-x}$K$_x$Fe$_2$As$_2$ were grown using self-flux method \cite{crystals,yliu2}.  Samples used for four-probe electrical resistivity measurements were cleaved from inner parts of large single crystals (with surface area up to 1 cm$^2$ and 0.3 mm thickness) and had dimensions of typically (2-3)$\times$0.5$\times$0.1 mm$^3$ with longer side along [100] tetragonal direction. Silver wires were soldered using Sn to the fresh-cleaved surface of the samples \cite{SUST,patent} to make electrical contacts with several microOhm resistance. Sample resistivity at room temperature, $\rho(300K)$, was doping independent within statistical error bars 
of geometric factor determination, $\pm$ 10\%. For all samples it was set to an average value as determined on a 
big array of crystals, $\rho(300K)$=300 $\mu \Omega$cm \cite{anisotropy}. Temperature dependent electrical resistivity, $\rho (T)$, measurements were performed down to 1.8~K in {\it Quantum Design} PPMS for sample screening. Measurements were performed in zero magnetic field. The sharpness of zero-field resistive transition was used as criterion for sample selection. The The composition of the selected samples was determined using electron probe microanalysis with wavelength dispersive spectroscopy (WDS). In Fig.~\ref{resistivity} we show low-temperature part of temperature-dependent resistivity of selected samples. Their position on doping phase diagram is indicated with triangles in the lower panel of Fig.~\ref{resistivity}.

Selected samples were glued with GE-varnish to a plastic platform, fitting single axis rotator of the 35~T DC magnet in National High Magnetic Field Laboratory in Tallahassee, Florida. Sample resistance was checked after mounting and found to be identical to the initial value. Sample long axis (current direction) was aligned by eye parallel to rotation axis (with accuracy of about 5$^\circ$). High-field measurements were made in He-cryostat with variable temperature control insert (VTI) allowing for temperatures down to 1.5~K.
The stepping motor driven rotator enabled {\it in situ} rotation with $\sim$0.1$^o$ resolution around a horizontal axis in a single axis rotation system of vertical 35~T magnetic field. In an ideal case of perfect parallel alignment of sample and rotation axes, during this rotation the direction of magnetic field with respect to the crystal traverses in the tetragonal (100) plane ($ac$-plane) always remaining perpendicular to the current. However, field-rotation plane may be somewhat inclined from (100) plane due to potential misalignment of sample and rotator axes, see \cite{JasonHc2} for details. This misalignment does not affect precision alignment in $H \parallel ab$ plane configuration ($\theta$=0), which was achieved by measuring angle dependent resistivity in a field slightly below the end of the resistive transition in field close to parallel to the plane configuration (see inset in the left top panel of Fig.~\ref{x0p17andx0p22}).

\section{Results}


In Fig.~\ref{x0p17andx0p22} we show isothermal magnetic field sweep resistivity data, $\rho(H)$, taken at different temperatures in magnetic fields aligned along $c$-axis ($\theta$=90$^o$, top panel) and precisely along the conducting plane ($\theta$=0$^o$, middle panel), for under-doped composition of Ba$_{1-x}$K$_x$Fe$_2$As$_2$ with $x$=0.17 (left column of panels). Inset in the top-left panel shows angular dependent resistivity in magnetic field slightly below $H_{c2}^{a}$ used for field alignment parallel to the conducting plane. Bottom-left panel summarizes $H-T$ phase diagrams as determined using transition mid-point criterion (symbols in top and middle panels). The use of this criterion is justified by small variation of the resistive transition width on application of magnetic field, and its independence on the extrapolation, typical problem for onset and offset criteria. Low value of $T_c$=12~K in the sample enables complete suppression of superconductivity at base temperature in $H \parallel c$ configuration and essential suppression in $H \parallel a$ configuration. The data in $H \parallel c$ configuration are in reasonable agreement with previous measurements in smaller field-temperature range \cite{ReidBaKHC2}, finding nearly $T$-linear $H_{c2}^{c}(T)$ without any sign of saturation on $T \to$0, contrary to WHH theory expectations \cite{WHH}. A clear tendency down-ward curvature with the tendency for saturation  is found in $H_{c2}^{a}(T)$ curve. 

In the right column top and middle panels of Fig.~\ref{x0p17andx0p22} we show raw resistivity field-sweep data in sample with $x$=0.22, the phase diagram is presented in the bottom panel. Magnetic field of 35~T $H \parallel c$ is sufficient to suppress superconductivity down to T=6~K ($T/T_c \approx$ 0.3), while in $H \parallel a$ superconductivity can be suppressed only down to 10~K ($T/T_c \approx$ 0.5). Despite limited range of magnetic field, the temperature-dependent anisotropic $H_{c2}(T)$ reveal the same trend as found in $x$=0.17 sample, with close to linear dependence and  small up-ward curvature in $H \parallel c$ and a tendency for saturation in $H \parallel a$.



\begin{figure}[tb]
\begin{center}
\includegraphics[width=0.90\linewidth]{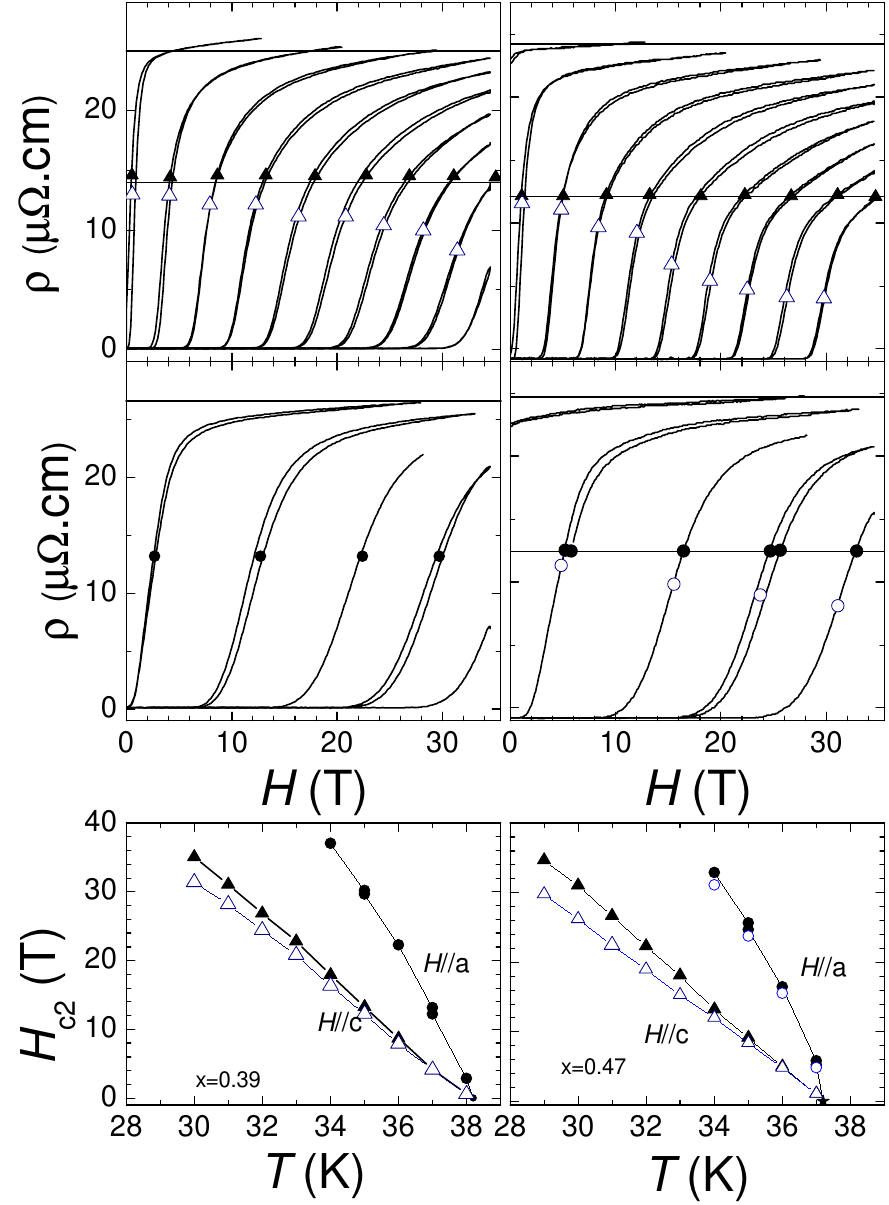}
\end{center}
\caption{(Left column of panels.) Magnetic field-dependent resistivity of sample Ba$_{1-x}$K$_x$Fe$_2$As$_2$ $x$=0.39 taken in isothermal conditions in magnetic fields oriented along tetragonal $c$-axis ($H \parallel c$, top panel, temperatures 38~K, 37~K, 36~K, 35~K, 34~K, 33~K, 32~K,31~K, 30~K, and 29~K left to right) and perpendicular to it ($H \parallel a$, mid panel, temperatures from left to right 38~K, 37~K, 36~K, 35~K, and 34~K). Bottom panel shows $H-T$ phase diagrams for two field orientations determined using mid-point criterion between up- and down- field sweeps sweeps (open symbols) and 
constant resistance criterion (line and solid symbols).  
(Right column of panels.) 
Magnetic field-dependent resistivity of sample Ba$_{1-x}$K$_x$Fe$_2$As$_2$ $x$=0.47 taken in isothermal conditions in magnetic fields oriented along tetragonal $c$-axis ($H \parallel c$, top panel, temperatures 38~K, 37~K, 36~K, 35~K, 34~K, 33~K, 32~K,31~K, 30~K, and 29~K left to right) and perpendicular to it ($H \parallel a$, mid panel, temperatures from left to right 38~K, 37~K, 36~K, 35~K, and 34~K). Bottom panel shows $H-T$ phase diagrams for two field orientations determined using mid-point criterion between up- and down- field sweeps (open symbols) and constant resistance criterion (lines and solid symbols).  
}%
\label{x0p39andx0p56}
\end{figure}

In Fig.\ref{x0p39andx0p56} we show raw resistivity data (top panels in $H \parallel c$, middle panels in $H \parallel a$ configurations) and $H-T$ phase diagrams (bottom panels) in samples close to optimal doping, $x$=0.39 (left column of panels) and $x$=0.47 (right column of panels).
Very narrow part of the phase diagram can be explored with 35~T magnetic field, however, even in this limited range the difference between close to $T$-linear $H_{c2}^{c}$ and down-curving
$H_{c2}^{a}$ is visible. Note the strong variation of the normal state resistivity with temperature in both samples, which makes impossible determination of $H_{c2}$ using same resistivity criterion. Transition midpoint is also ill-defined criterion for the sample, because of rounding of $\rho (H)$ curves in the normal state, presumably due to filamentary superconductivity in the normal state.


\begin{figure}[tb]
\begin{center}
\includegraphics[width=0.90\linewidth]{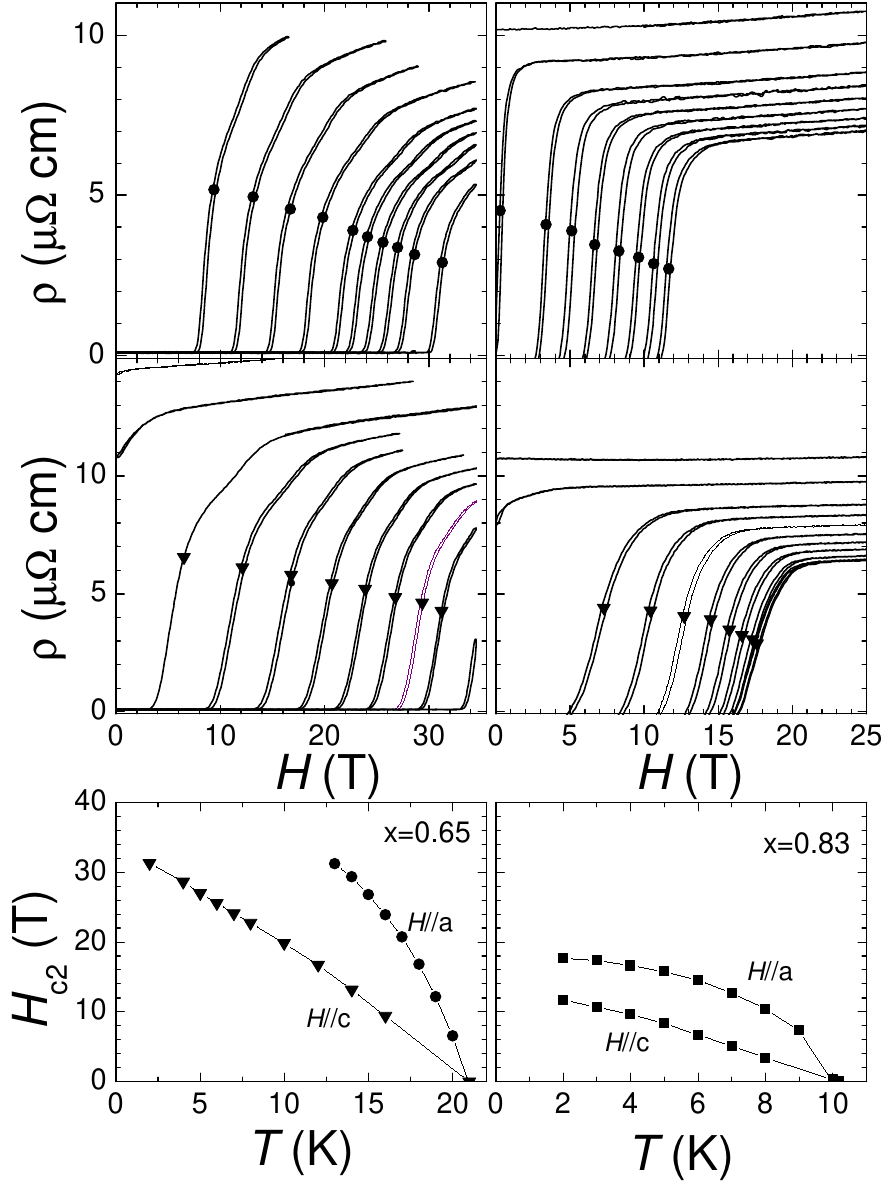}
\end{center}
\caption{(Left column of panels.) Magnetic field-dependent resistivity of sample Ba$_{1-x}$K$_x$Fe$_2$As$_2$ $x$=0.65 taken in isothermal conditions in magnetic fields oriented along tetragonal $c$-axis ($H \parallel c$, top panel, temperatures 16~K, 14~K, 12~K, 10~K, 8~K, 7~K, 6~K, 5~K, 4~K, and 2~K left to right) and parallel to the plane ($H \parallel a$, mid panel, temperatures from left to right 24~K, 22~K, 20~K, 19~K, 18~K, 17~K, 16~K, 15~K, 14~K, 13~K, and 11~K). Bottom panel shows $H-T$ phase diagrams for two field orientations determined using mid-point criterion between up- and down- field sweeps.   
(Right column of panels.)
Magnetic field-dependent resistivity of sample Ba$_{1-x}$K$_x$Fe$_2$As$_2$ $x$=0.83 taken in isothermal conditions in magnetic fields oriented along tetragonal $c$-axis ($H \parallel c$, top panel, temperatures 12~K, 10~K, 8~K, 6~K, 5~K, 4~K, 3~K,and 2~K left to right) and perpendicular to it ($H \parallel a$, mid panel, temperatures from left to right 11~K, 9~K, 8~K, 7~K, 6~K, 5~K, 4~K, 3~K and 2~K). Bottom panel shows $H-T$ phase diagrams for two field orientations determined using mid-point criterion between up- and down- field sweeps.   
}%
\label{x0p65andx0p83}
\end{figure}

In Fig.~\ref{x0p65andx0p83} we show resistivity vs. field curves in configurations $H \parallel c $ (top panels) and $H \parallel a$ (middle panels)  for overdoped samples $x$ =0.65 (left column of panels) and $x$=0.83 (right column of panels). Smaller value of $T_c$ enables characterization of the whole phase diagram for 10~K class sample $x$=0.83. Note decrease of the normal state resistivity on cooling in both compositions, the tendency for $H_{c2}^{c}(T)$ saturation on cooling and pronounced tendency for saturation at temperatures close to zero-field $T_c$ in $H \parallel a$ configuration. 

\subsection{Experimental summary}

\begin{figure}[tb]
\begin{center}
\includegraphics[width=0.90\linewidth]{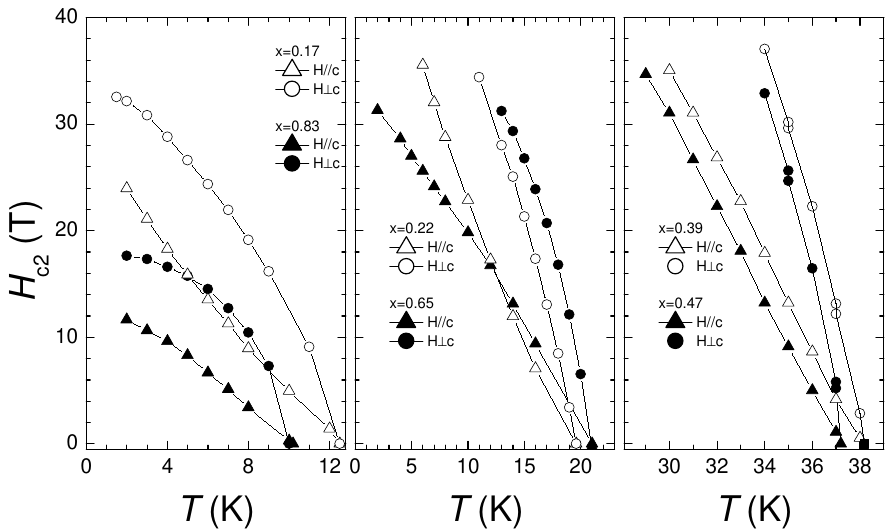}
\end{center}
\caption{(Color Online) Left panel. Comparison of the $H-T$ phase diagrams of "10~K" pair of underdoped ($x$=0.17) and overdoped ($x$=0.83) samples. Middle panel shows similar comparison for a "20~K" pair of samples, underdoped ($x$=0.22) and overdoped ($x$=0.65), right panel is data for samples close to optimum doping $x$=0.39 and $x$=0.47. Note clear tendency to saturation in both overdoped compositions for magnetic field configuration $H \parallel a$. 
}%
\label{phased_10Kand20Kcomparison}
\end{figure}

In Fig.~\ref{phased_10Kand20Kcomparison} we make direct comparison of the $H-T$ phase diagrams of 10~K class samples (under-doped $x$=0.17 and over-doped $x$=0.83) (top panel), 20~K class samples (under-doped $x$=0.22 and over-doped $x$=0.65) (middle panel) and of optimally doped 38~K samples, $x$=0.39 and $x$=0.47 (bottom panel). This comparison highlights the difference between two doping regimes. The  
$H_{c2}^{c}(T)$ shows small upward curvature in underdoped compositions, somewhat reminiscent of the dependence in layered superconductors \cite{Andy}, and in multi-band superconductors \cite{KoganProzorov}. Slight tendency for saturation of $H_{c2}^{c}(T)$ may be found in overdoped compositions. A tendency for $H_{c2}(T)$ saturation for orbital limiting is expected to become visible below $T/T_c<$0.3 \cite{WHH}, see Fig.~\ref{Hc2slopeadjusted} below.
Deviation from this prediction in iron-based superconductors was discussed in multi-band scenario \cite{KoganProzorov,JasonHc2}. Indeed heat capacity \cite{HardyBaK} and London penetration depth studies \cite{KChoSA} suggest pronounced multi-band effects, with the gap magnitude on different sheets of the Fermi surface varying by a factor of approximately two. 

\begin{figure}[tb]
\begin{center}
\includegraphics[width=0.90\linewidth]{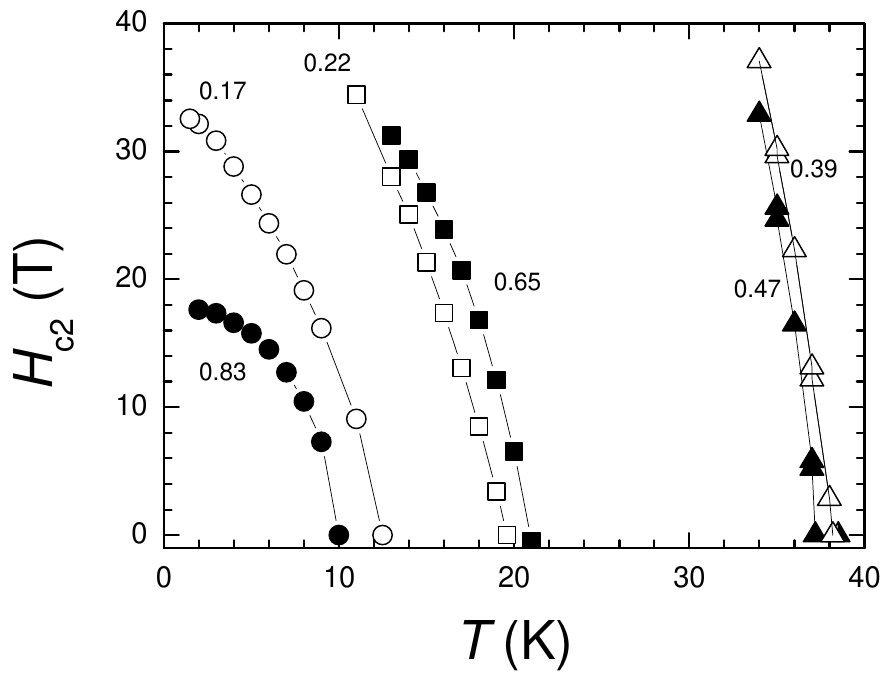}
\end{center}
\caption{(Color Online) Comparison of the $H-T$ phase diagrams in parallel magnetic field $H_{c2}^{a}(T)$ for samples of Ba$_{1-x}$K$_x$Fe$_2$As$_2$, for compositions (left to right) with $x$=0.83(solid circles), $x$=0.17 (open circles), $x$=0.22 (open squares), $x$=0.65 (solid squares), $x$=0.39 (open triangles) and $x$=0.47 (solid triangles). 
}%
\label{phased_parallel}
\end{figure}

The dependence in the configuration with magnetic field parallel to the plane is even more intriguing.
  In Fig.~\ref{phased_parallel} we make comparison of the data for all compositions in precision aligned $H \parallel a$ conditions. Note much more pronounced curvature of $H_{c2}^{a}(T)$ close to zero field $T_c$ in the overdoped compositions. 

\section{Discussion}

\begin{figure}[tb]
\begin{center}
\includegraphics[width=0.90\linewidth]{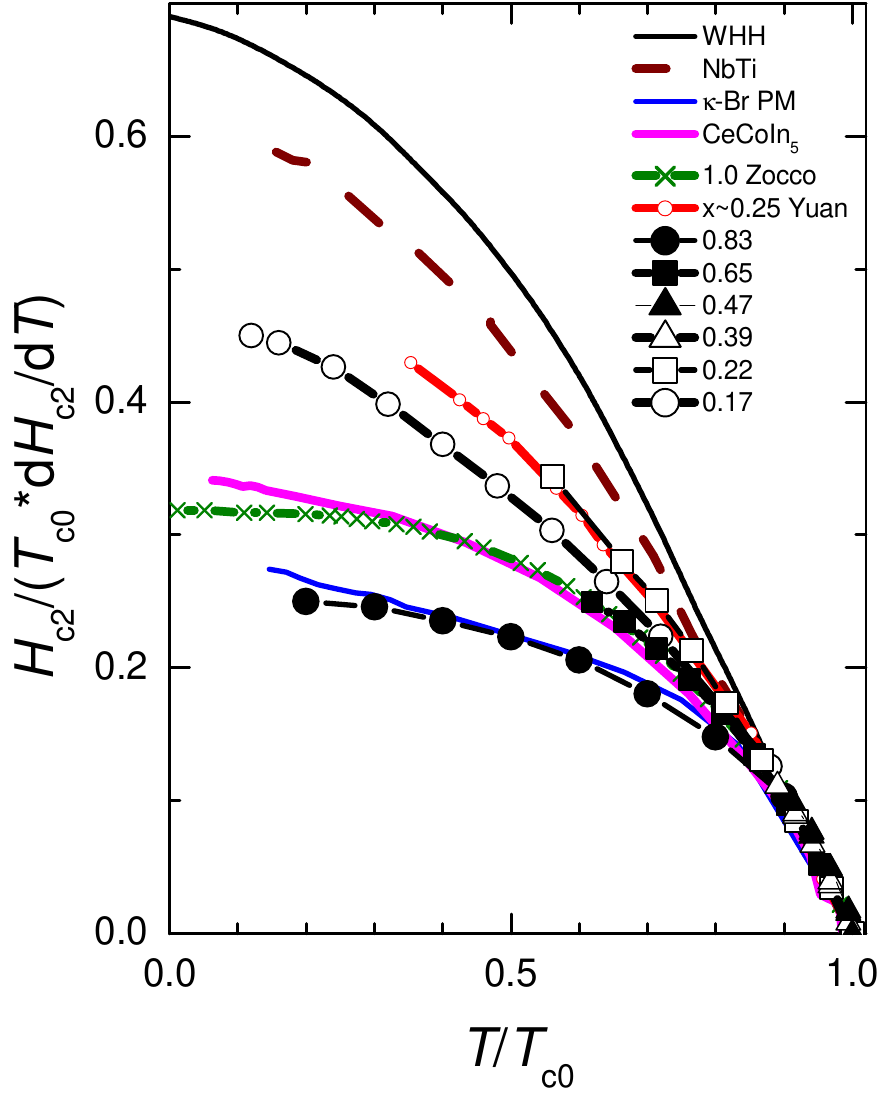}
\end{center}
\caption{(Color Online) Comparison of the $H-T$ phase diagrams of Ba$_{1-x}$K$_x$Fe$_2$As$_2$ in magnetic field precisely parallel to the plane $H_{c2}^{a}(T)$ using normalized temperature $T/T_{c0}$ and magnetic field $H/(T_{c0}dH_{c2}(T)/dT$) scales. For reference we show expectations for orbital limiting mechanism in WHH theory \cite{WHH} (black solid line), experimentally determined $H_{c2}$ line for conventional isotropic superconductor NbTi with dominant orbital limiting (dashed brown line), layered organic superconductor $\kappa$-(BEDT-TTF)$_2$Cu[N(CN$_2$]Br ($\kappa$-Br) in magnetic field parallel to conducting plane in which paramagnetic limiting starts in the very vicinity of $T_c$ \cite{Kovalev} (blue line) and paramagnetically limited CeCoIn$_5$ \cite{Bianchi} (magenta line). We also plot data for KFe$_2$As$_2$ as determined from magnetostriction and thermal expansion measurements in $x$=1 by Zocco {\it et al.} \cite{Zocco} (green crosses and line) and from resistivity measurements in pulsed field for sample with $T_c \approx$28~K corresponding roughly to $x$=0.25 from Yuan {\it et al.} (red line with open triangles) \cite{Singleton}.  Solid symbols represent overdoped compositions with $x$=0.83(circles), $x$=0.65 (squares) and $x$=0.47 (triangles). Open symbols are used for under-doped compositions $x$=0.17 (circles), $x$=0.22 (squares) and for optimally doped  $x$=0.39 (triangles).
}%
\label{Hc2slopeadjusted}
\end{figure}

There are two mechanisms that determine the upper critical field of superconductors. The first one, determined by the supercurrent flow to screen the magnetic field, is referred to as orbital limiting and described by WHH theory \cite{WHH}. The upper critical field at $T \to 0$ limit, $H_{c2}(0)$, in WHH theory is determined by the slope of the $H_{c2}(T)$ curve close to $T_c$, and as $T$ goes to zero the curve shows downward deviation from linear dependence and eventual saturation towards the value $H_{c2}(0) \approx 0.7T_c \frac{dH_{c2}}{dT}$ in isotropic case. In Fig.~\ref{Hc2slopeadjusted} we show the temperature dependent $H_{c2}(T)$ as expected in WHH theory, and the data for isotropic NbTi, as a typical experimental observation. 

Rather rare exceptions, when the upper critical field is not determined by the orbital limiting, are found in the materials in which orbital motion of electrons is hampered by either short mean free path, heavy mass of conduction electrons in heavy fermion materials or weak links between the conducting layers in Josephson structures or in naturally highly electronically anisotropic layered materials \cite{Ishiguro}, provided that the magnetic field is aligned precisely parallel to the conducting layer. In this situation the upper critical field $H_{c2}$ is determined by Zeeman splitting of electron levels, known as Clogston-Chandrasekhar \cite{CC} paramagnetic limit. This field is determined by a decrease of paramagnetic energy becoming equal to condensation energy of superconductor. In weak coupling BCS superconductors the paramagnetic limiting field is determined in $T \to 0$ limit as $H_p$=1.8$T_c$, where $H_p$ is field in Tesla and $T_c$ is in Kelvin. 
Note, however, that even in materials with dominant paramagnetic effects, the behavior of $H_{c2}(T)$ line close to zero field $T_c$ is always determined by the orbital limiting mechanism, so that the slope of $H_{c2}$ lines at $T_c$ is reflecting anisotropy of the electronic structure. 
The width of the temperature range in which orbital limiting mechanism is dominant depends on the ratio of orbital and paramagnetic limiting fields (Maki parameter) \cite{Makiparameter}. In strongly anisotropic materials, like organic $\kappa$-(BEDT-TTF)$_2$[Cu[N(CN)$_2$]Br ($T_c$=12.8~K, $\gamma_{\rho} \sim $10$^4$), where BEDT-TTF stands for bis(ethylenedithio)tetrathiafulvalene, this range is confined to the very vicinity of $T_c$ \cite{Kovalev}  and the experimentally determined $H_{c2}(T)$ curve makes a good experimental example for the shape of the upper critical field in paramagnetically limited superconductors \cite{Ishiguro}. This dependence is shown in Fig.~\ref{Hc2slopeadjusted} with blue line. For all the curves in the figure the slope of lines near $T_c$ was adjusted to match orbital limiting expectations.

In Fig.~\ref{Hc2slopeadjusted} we plot the data for the 10~K and 20~K class samples of Ba$_{1-x}$K$_x$Fe$_2$As$_2$ in $H \parallel a$ configuration in comparison with curves for orbital and paramagnetic limiting cases. Note that the curve for $x$=0.83 is closely following expectations for paramagnetic limiting, with negligible orbital contribution. Interestingly, even the value of the upper critical field for this sample is close to expectations for weak coupling superconductor), $H_p$=1.8*10=18~T (see Fig.~\ref{phased_parallel}). The curves for other overdoped composition $x$=0.65 still is very close to expectations for paramagnetic limiting, and closely follows experimental data for CeCoIn$_5$ (magenta line) strongly paramagnetically limited superconductor \cite{Bianchi}. The curves for optimally doped samples $x$=0.39 (open up-triangles in Fig.~\ref{Hc2slopeadjusted}) and $x$=0.47 (solid up-triangles) are defined in a too narrow range to distinguish between orbital and paramagnetic limits clearly. The data for underdoped compositions $x$=0.17 (open circles) and $x$=0.22 (open squares) are close to orbital limiting. For reference we also plot literature data for the sample $x$=1 \cite{Zocco} and for the underdoped sample with $T_c\approx$28~K ($x \sim$0.25) \cite{Singleton}. The curve for the latter sample is very close to the dependence for $x$=0.22 in our measurements, the curve for $x$=1 is somewhat more deviating from the closest data for $x$=0.83 in our measurements, revealing somewhat higher saturation value of $H_{c2}^{a}$.  

The findings of our study are somewhat unexpected. Paramagnetic effects in $H_{c2}(T)$ are usually found is quite anisotropic materials, like layered organic $\kappa$-(BEDT-TTF)$_2$Cu[N(CN)$_2$]Br, with $\gamma_{\rho} \sim 10^3 -10^4$. The anisotropy of the resistivity and upper critical fields near $T_c$ in Ba$_{1-x}$K$_x$Fe$_2$As$_2$ is significantly lower \cite{YLiu1}, in KFe$_2$As$_2$ for example, $\gamma_H \sim$5 to 7 \cite{TerashimaHc2,YLiu1,Zocco} and $\gamma _{\rho} \sim$30 \cite{ReidK}. This observation might suggest that increased effective masses, as found in heat capacity \cite{Zocco,Zocco2,HardyBaK,Stewart} and transport \cite{SLi,Valentin} studies, are pushing orbital upper critical fields up for both in-plane and out-of-plane field orientations, similar to the heavy fermion materials.  In situation like this, asymmetry of the doping evolution of $H_{c2}{^a}$ may be a reflection of increasing effective mass, as reflected in increasing electronic heat capacity coefficient $\gamma_N \equiv C_{el}/T$ on approaching quantum critical point in the overdoped compositions, rather than electronic anisotropy. This would suggest that $H_{c2}^{a}/T_c$ ratio should be increased on the overdoped side, which is not the case.

On the other hand, the upper critical field in $H \parallel ab$ configuration in the underdoped compositions is intermediate between paramagnetic and orbital limiting, despite similar anisotropy values of $\gamma_H$ and significantly lower normal state electronic heat capacity $\gamma_N$ \cite{HardyBaK}. Notable asymmetry may be related to asymmetry found in the shape of $H_{c2}^{c}(T)$ with upward deviations from linear dependence in the under-doped $x$=0.17 compared to down-ward in overdoped $x$=0.83 and $x$=0.65. In multi-band scenario upward deviation may suggest notable increase in effective gap magnitude in $x$=0.17 on cooling, which can lead to upward deviation even in paramagnetic limiting regime. It is known that multi-band effects are needed to account for heat capacity \cite{HardyBaK} and London penetration depth \cite{KChoSA} for all compositions of Ba$_{1-x}$K$_x$Fe$_2$As$_2$. Our observation may be suggestive of a stronger multi-band effects on the under-doped side. 

Another possible source of asymmetry may be coexistence with long-range magnetic order in the compositions $x$=0.17 and $x$=0.22 on under-doped side of the phase diagram, but not in $x$=0.65 and $x$=0.83 on the over-doped side. Long range magnetic order is known to affect the nodal structure of the superconducting gap, as suggested theoretically \cite{Chubukov} and observed experimentally in London penetration depth \cite{HyunsooBaK} and thermal conductivity \cite{ReidBaKHC2} measurements. To the best of the author knowledge, the effect of the coexisting magnetism on the upper critical field $H_{c2}$ was not discussed for iron-based superconductors.

\section{Conclusions}

By performing study of the anisotropic upper critical fields in Ba$_{1-x}$K$_x$Fe$_2$As$_2$ series of compounds for a range of compositions spanning from underdoped ($x$=0.17) to overdoped ($x$=0.83) in precision alignment magnetic field orientation conditions, we find strong paramagnetic limiting of $H_{c2}^{a}$ in the overdoped compositions, but not in the underdoped compositions. We speculate that much more pronounced multi-band effects on the under-doped side may be responsible for the asymmetry, as suggested by the difference in the shapes of $H_{c2}^{c}(T)$ dependences.

\section{Acknowledgements}

We thank V.~G.~Kogan for critical reading of the manuscript and useful discussions.
The experimental work was supported by the U.S. Department of Energy
(DOE), Office of Basic Energy Sciences, Division of Materials Sciences
and Engineering. The experimental research was performed at Ames Laboratory, 
which is operated for the U.S. DOE by Iowa State University under
Contract No.~DE-AC02-07CH11358. 
Work at the National High Magnetic Field Laboratory is supported by the NSF Cooperative Agreement No. DMR 1157490 and by the State of Florida. 


\end{document}